\documentclass[EPiCempty]{easychair}

\usepackage{doc}
\usepackage{amsmath,amssymb,amsfonts}
\usepackage{todonotes}
\usepackage[font=small,labelfont=bf]{caption}
\usepackage[binary-units,per-mode=symbol]{siunitx}
\usepackage{tikz}
\usetikzlibrary{patterns}
\usepackage{pgfplots}
\pgfplotsset{compat=newest}
\usepgfplotslibrary{units}
\makeatletter
\pgfplotsset{
 unit code/.code 2 args=
   \begingroup
   \protected@edef\x{\endgroup\si{#2}}\x
}
\makeatother
\usetikzlibrary{pgfplots.groupplots}
\definecolor{lightcorered}{RGB}{250,150,150}
\definecolor{corered}{RGB}{200,0,0}
\definecolor{coredarkgray}{RGB}{51,51,51}
\definecolor{coreblue}{RGB}{0, 46, 125}
\usepackage[nohyperlinks,nolist]{acronym} 
\usepackage[absolute,showboxes]{textpos}

\title{SDN4CoRE: A Simulation Model for Software-Defined Networking for Communication over Real-Time Ethernet}

\author{
Timo H\"ackel \and Philipp Meyer \and Franz Korf \and Thomas C. Schmidt
}

\institute{
 \href{http://www.haw-hamburg.de/ti-i}{\textit{Dept. Computer Science}},
\href{http://www.haw-hamburg.de/ti-i}{\textit{Hamburg University of Applied Sciences}}, Germany \\
\{\href{mailto:timo.haeckel@haw-hamburg.de}{timo.haeckel}, \href{mailto:philipp.meyer@haw-hamburg.de}{philipp.meyer}, \href{mailto:franz.korf@haw-hamburg.de}{franz.korf}, \href{mailto:t.schmidt@haw-hamburg.de}{t.schmidt}\}@haw-hamburg.de
}

\authorrunning{H\"ackel, Meyer, Korf and Schmidt}

\titlerunning{SDN4CoRE Simulation Model}

\begin{document}

\maketitle

\setlength{\TPHorizModule}{\paperwidth}
\setlength{\TPVertModule}{\paperheight}
\TPMargin{5pt}
\begin{textblock}{0.8}(0.1,0.02)
     \noindent
     \footnotesize
     If you cite this paper, please use the original reference:
     T. H{\"a}ckel, P. Meyer, F. Korf, and T.~C. Schmidt. SDN4CoRE: A Simulation Model for Software-Defined Networking for Communication over Real-Time Ethernet. In: \emph{Proceedings of the 6th International OMNeT++ Community Summit}. September, 2019, Easychair.
\end{textblock}

\begin{abstract}
Ethernet has become the next standard for automotive and industrial automation networks.
Standard extensions such as IEEE 802.1Q Time-Sensitive Networking (TSN) have been proven to meet the real-time and robustness requirements of these environments. 
Augmenting the TSN switching by Software-Defined Networking functions promises additional benefits:      
A programming option for TSN devices can add much value to the resilience, security, and adaptivity of the environment.
Network simulation allows to model highly complex networks before assembly and is an essential process for the design and validation of future networks. 
Still, a simulation environment that supports programmable real-time networks is missing.
    
This paper fills the gap by sharing our simulation model for Software-Defined Networking for Communication over Real-Time Ethernet (SDN4CoRE) and present initial results in modeling programmable real-time networks. 
In a case study, we show that SDN4CoRE can simulate complex programmable real-time networks and allows for testing and verifying the programming of real-time devices.
\end{abstract}



%
%



\begin{acronym}
	\acro{API}[API]{Application Programming Interface}
	\acro{AVB}[AVB]{Audio Video Bridging}
	\acro{ARP}[ARP]{Address Resolution Protocol}
	\acro{BE}[BE]{Best-Effort}
	\acro{CAN}[CAN]{Controller Area Network}
	\acro{CBM}[CBM]{Credit Based Metering}
	\acro{CBS}[CBS]{Credit Based Shaping}
	\acro{CMI}[CMI]{Class Measurement Interval}
	\acro{CoRE}[CoRE]{Communication over Realtime Ethernet}
	\acro{CT}[CT]{Cross-Traffic}
	\acro{DoS}[DoS]{Denial of Service}
	\acro{DPI}[DPI]{Deep Packet Inspection}
	\acro{ECU}[ECU]{Electronic Control Unit}
	\acroplural{ECU}[ECUs]{Electronic Control Units}
	\acro{GCL}[GCL]{Gate Control List}
	\acro{HTTP}[HTTP]{Hypertext Transfer Protocol}
	\acro{HMI}[HMI]{Human-Machine Interface}
	\acro{IA}[IA]{Industrial Automation}
	\acro{IEEE}[IEEE]{Institute of Electrical and Electronics Engineers}
	\acro{IoT}[IoT]{Internet of Things}
	\acro{IP}[IP]{Internet Protocol}
	\acro{ICT}[ICT]{Information and Communication Technology}
	\acro{LIN}[LIN]{Local Interconnect Network}
	\acro{MOST}[MOST]{Media Oriented System Transport}
	\acro{OEM}[OEM]{Original Equipment Manufacturer}
	\acro{RC}[RC]{Rate-Constrained}
	\acro{REST}[ReST]{Representational State Transfer}
	\acro{RPC}[RPC]{Remote Procedure Call}
	\acro{SDN}[SDN]{Software-Defined Networking}
	\acro{SDN4CoRE}[SDN4CoRE]{Software-Defined Networking for Communication over Real-Time Ethernet}
	\acro{SOA}[SOA]{Service-Oriented Architecture}
	\acro{SOME/IP}[SOME/IP]{Scalable service-Oriented MiddlewarE over IP}
	\acro{SR}[SR]{Stream Reservation}
	\acro{SRP}[SRP]{Stream Reservation Protocol}
	\acro{TCP}[TCP]{Transmission Control Protocol}
	\acro{TDMA}[TDMA]{Time Division Multiple Access}
	\acro{TSN}[TSN]{Time-Sensitive Networking}
	\acro{TSSDN}[TSSDN]{Time-Sensitive Software-Defined Networking}
	\acro{TT}[TT]{Time-Triggered}
	\acro{TTE}[TTE]{Time-Triggered Ethernet}
	\acro{UDP}[UDP]{User Datagram Protocol}
	\acro{QoS}[QoS]{Quality-of-Service}
	\acro{WS}[WS]{Web Services}
\end{acronym}

\vspace{-5pt}
\section{Introduction} 
\label{sec:introduction}
In recent years, Ethernet has emerged to become the next standard for automotive networks.
Complementary protocols such as IEEE 802.1Q \ac{TSN} have proven to meet the real-time and robustness requirements of these environments.
On the other hand, the \ac{SDN} paradigm has revolutionized campus and data center networks.
Separating the control from the data plane of network devices at a central control unit with global network knowledge enables simple and fast-forwarding at devices, while high-level control applications can steer the entire network.
 \ac{SDN}, though, cannot grant \ac{QoS} guarantees to the forwarding plane.
First results from integrating the \ac{SDN} paradigm with \ac{TSN} standards are promising \cite{hmks-snsti-19}. 
A programming option for \ac{TSN} devices can add much value to the resilience, security, and adaptivity of the environment.
Although the application of the combination of \ac{TSN} and \ac{SDN} is expected to expand with the introduction of 5G and the Internet of Things, we focus on the use case in automotive networks in this work.

Network simulation allows highly complex networks to be modeled before the assembly. 
This is an important technique for the design and validation of future networks and therefore remains an active research topic.
The discrete event simulation platform OMNeT++ suits well as a simulation toolchain for automotive communication, as shown in prior work~\cite{sdks-eoifs-11}.

In this paper, we present our simulation model for \ac{SDN4CoRE} and describe the first results in modeling programmable real-time networks. 
\ac{SDN4CoRE} is built on top of the INET framework and uses the CoRE4INET simulation models developed in previous work~\cite{mkss-smcin-19}. 
To make the network devices programmable, we implement client and server modules for the NetConf protocol and integrate the OpenFlowOMNeTSuite~\cite{kj-oeoJR-13} for OpenFlow protocol modules.
In our case study, we show how our model can be used to test and evaluate configuration mechanisms in real-time networks. 

The remainder of this paper is structured as follows. 
Section~\ref{sec:background_&_related_work} provides background knowledge and related work. 
The concept of programming options for the different real-time Ethernet components is discussed in Section~\ref{sec:make_communication_over_real_time_ethernet_programmable}.
In Section~\ref{sec:simulation_model_for_tssdn}, we describe the components of the \ac{SDN4CoRE} simulation models, followed by a case study in Section~\ref{sec:case_study}. 
Finally, Section~\ref{sec:conclusion_&_future_work} concludes this work with an outlook on future work.


\vspace{-5pt}
\section{Background and Related Work} 
\label{sec:background_&_related_work}
Today, more and more software components get deployed in automotive networks, demanding a steady increase in communication bandwidth and timing guarantees.
Ethernet has emerged as the next high-bandwidth communication technology for in-car networks.
There have been several attempts to introduce support for real-time requirements in Ethernet networks.
Time-Triggered Ethernet (AS6802) provides a synchronous \ac{TDMA} implementation, as well as rate-constrained traffic classes for Ethernet. 
\ac{AVB} defines dynamic bandwidth reservation mechanisms for streams and traffic shapers to guarantee maximum latency. 
\ac{AVB}s successor \acl{TSN} (IEEE 802.1Q-2018~\cite{ieee8021q-18}) is a set of standards which are defined by the \ac{TSN} task group of the IEEE. 
These standards extend Ethernet to concurrently forward real-time and \ac{CT}. 
It supports both synchronous \ac{TDMA} traffic and asynchronous bandwidth reservation for streams.
CoRE4INET~\cite{mkss-smcin-19} (Communication over  Real-time Ethernet for INET) is a suite of real-time Ethernet simulation models developed over the past decade.
Currently, CoRE4INET supports the AS6802 protocol suite, traffic shapers of  Ethernet \ac{AVB}, and implementations of IEEE 802.1Q, as well as models to map IP traffic to real-time traffic classes. 

As communication requirements increase, so does the complexity and network adaptability requirements.
The \ac{SDN} paradigm promises to solve this problem by moving the control logic out of the network devices into a central control unit~\cite{mabpp-oeiJR-08}.
Kreutz et al. discuss the paradigms and concepts of \ac{SDN} in their comprehensive survey~\cite{krvra-sdnJR-15}. 
The network logic is split into three layers: 
(1) The data plane on which each switch forwards packets according to the flow rules,
(2) the control plane on which each switch is connected to a logically (not necessarily physically) centralized controller that manages the forwarding logic, and 
(3) the management plane on which network administrators manage the controller applications. 
The communication between the \ac{SDN} controller and the switches is specified in the OpenFlow standard of the ONF~\cite{o-ossJR-15}. 
The OpenFlowOmnetSuite -- originally developed at the University of W\"urzburg -- is an OMNeT++ simulation model that implements a concept of \ac{SDN} with OpenFlow~\cite{kj-oeoJR-13}. 
It provides the OpenFlow standard message types, implementations of forwarding devices, a \ac{SDN} controller implementation, and interfaces for controller applications. 

In \ac{TSN}, configuration of various components is enabled through the NetConf protocol (RFC~6241~\cite{RFC-6241}).
It specifies a management architecture in which the managed unit contains a NetConf server, and the administrative unit connects through a NetConf client.
On the server-side, configurations are stored in configuration data stores. 
NetConf provides \ac{RPC} operations such as \textit{get-config} and \textit{edit-config}, which allow the client to request or edit the configuration data stores at the server. 
\ac{SDN4CoRE} provides a base implementation of the NetConf protocol, including client, server, data stores, and operations. 

Rather little work has been done to combine the concepts of \ac{TSN} and \ac{SDN}. 
Nayak et al. mention the term "Time-Sensitive Software-Defined Network" in 2016 for the first time and show a concept of a programmable scheduler~\cite{ndr-tssJR-16}. 
In previous work, we introduced a concept of software-defined networks supporting time-sensitive in-vehicle communication~\cite{hmks-snsti-19} along with a switching methodology for \ac{TSN} streams, the programming of the bandwidth reservation via the \ac{SRP}, and a detailed analysis.

To the best of our knowledge, no simulation environment supports programmable real-time networks.
Making the modules of CoRE4INET programmable allows using other available simulation models as well, such as CAN bus signals and gateways to model realistic in-vehicular communication~\cite{mkss-smcin-19}.


\vspace{-5pt}
\section{Programmable Real-Time Ethernet Communication } 
\label{sec:make_communication_over_real_time_ethernet_programmable}
Introducing a programming option for real-time Ethernet devices that improves resilience, security, and adaptability of the environment requires changes on all three layers of the \ac{SDN} concept:
    Forwarding devices must provide a programming interface using open standards,
    control functionality for real-time must be extracted from the switches and integrated into the \ac{SDN} controller, and
    controller applications must be able to program and manage real-time devices.

\begin{figure}
        \centering
        \includegraphics[width=.7\linewidth, trim= 1.55cm .5cm 1.55cm .5cm, clip=true]{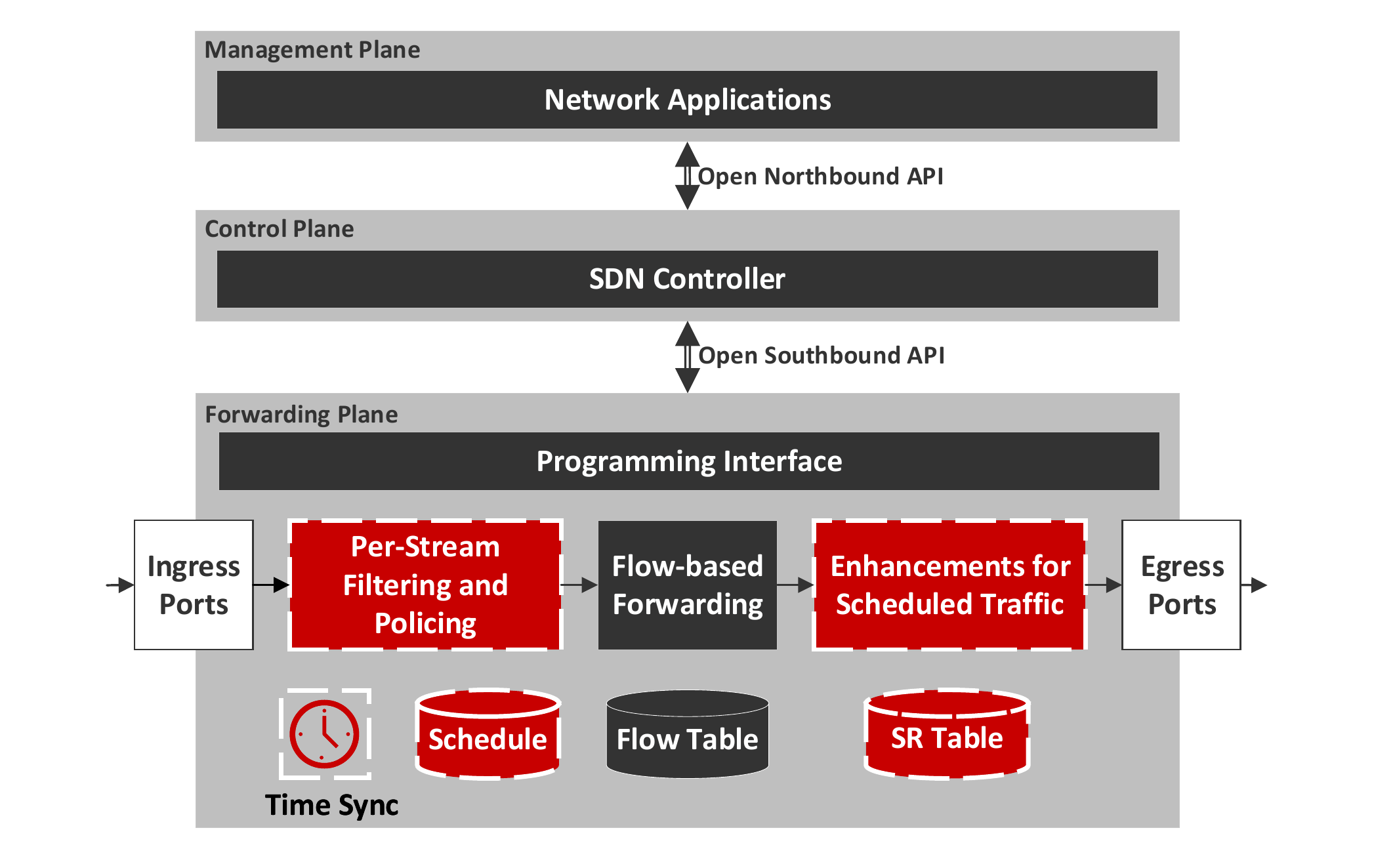}
        \includegraphics[width=.4\linewidth, trim= .7cm .6cm .7cm .3cm, clip=true]{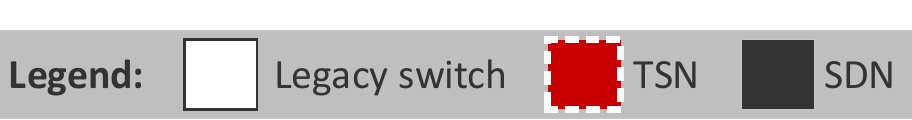}
        \caption{Components of a programmable real-time switch.}
        \label{fig:switches}
        \vspace{-15pt}
    \end{figure}

In real-time Ethernet and \ac{SDN}, switches contain additional modules to extend the functionality of regular switching hardware, which is depicted in Figure~\ref{fig:switches}.
One of the additional modules introduced in some real-time communication is an ingress control module. 
In \ac{TSN} this is the ``Per-Stream Filtering and Policing'' module.
It is used to filter incoming Ethernet frames for controlling bandwidth and arrival times.
A similar module is used at the egress of a switch and implements in particular priority queuing, real-time scheduling and traffic shaping.
In \ac{TSN} this module is called ``Enhancements for Scheduled Traffic'' and specified in IEEE 802.1Qbv.  
It uses a \ac{GCL} to indicate, which 802.1Q priorities are allowed to pass through a particular port at a specific time.
The scheduling information is stored in the ``Schedule'' table and needs a precisely synchronized time at all network devices, which is managed by the ``Time Sync'' module.
For stream-based bandwidth reservation, the ``SR Table'' module contains all registered talkers and listeners for time-sensitive streams.

In a \ac{SDN} switch, the forwarding module of a standard Ethernet switch is replaced by a flow-based forwarding module that performs flow table lookups based on packet match rules.
The \ac{SDN} controller performs tasks such as topology discovery, MAC-Address learning, and route determination.
The programming interface between the switch and the \ac{SDN} controller is an open southbound API that implements standard protocols such as NetConf or OpenFlow.
Additional network applications can be executed on top of the controller.

In programmable real-time networks, flow-based operations require merging the modules of the real-time and the \ac{SDN} switching components.
To ensure that the real-time capabilities are not altered in any way, the ingress and egress control modules must remain unchanged by the additional programming option.
When data packets arrive, the ingress control manages the timing and applies stream-based filters.
After that, the packet is matched against the flow table, and the discovered actions are executed.
The packet then gets forwarded to the correct egress ports, where the egress control manages the timing and shaping of the outgoing traffic. 
If no corresponding forwarding rule exists, the packet is dropped by default, while most controller applications insert the default rule to forward the packet to the controller.

\ac{SDN} controller applications manage the forwarding and store information for the control functionality of the \ac{SDN} devices.
This usually includes features such as neighbor discovery or source MAC address learning.  
Because many real-time Ethernet protocols use multicast MAC addresses to transfer data to a group of receivers, the controller applications must be able to learn and control multicast MAC addresses. 
\ac{AVB} and \ac{TSN} use the \ac{SRP} to announce the sender (talker) and receivers (listener) of the multicast groups.
Therefore, the controller application must receive and understand the \ac{SRP} packets.
When a forwarding device receives a \ac{SRP} message, it is forwarded directly to the \ac{SDN} controller via the OpenFlow protocol.
The controller application then registers the talker or listener in its SR table and sends the \ac{SRP} message back to the switch via the OpenFlow protocol. 
The switch updates its SR table and forwards the message to the next hop. 
In this way, the `talker advertise' and `listener ready' messages are spread across the network, and each switch goes through the same process until the messages reach the clients. 
When a new client subscribes to a stream with a `listener ready' message, a forwarding rule for the \ac{TSN} stream is inserted in the switches flow table before the `listener ready' command is sent back to the switch and forwarded along the path. 
In a future release, the modules handling the \ac{SRP} protocol could be removed from the switches to simplify them.
The controller could insert a match rule for the \ac{SRP} messages to receive them directly. 
Subsequently, the \ac{SRP} tables could then be updated directly via the NetConf protocol.

Besides the dynamic stream reservation, the controller must be able to control the scheduled traffic.
Therefore, the switch must provide a specific NetConf data store to control the scheduling tables.
In this work, we implemented such a data store for the gate control of 802.1Qbv. 
To change the \ac{GCL}, the controller sends an `edit config' message containing updates or a full configuration for the active \ac{GCL}. 
The NetConf server module forwards those messages to the data store which then updates the \ac{GCL} in the corresponding port.
In a future release, additional modules may be added to the data store.
For example, the controller could program the 802.1Qci filters, gates, and meters.
On the other hand, algorithms to calculate a new schedule for the entire network could be implemented to adapt the schedules dynamically.


\vspace{-5pt}
\section{SDN4CoRE Simulation Model} 
\label{sec:simulation_model_for_tssdn}
The \ac{SDN4CoRE} simulation module is based on the discrete event simulator OMNeT++ (\url{https://omnetpp.org/}) and the INET framework (\url{https://inet.omnetpp.org/}).
It uses the CoRE4INET~\cite{smbk-eotpd-16} simulation model for real-time Ethernet communication and extends some modules to be programmable.
On the other hand, \ac{SDN4CoRE} uses the OpenFlow protocol implemented in the OpenFlowOMNeTSuite~\cite{kj-oeoJR-13}. 
We forked and extended this suite to harmonize with our models. 
For example, we introduced interface modules to enhance flexibility and modularity and created a new flow table structure.


\begin{figure}
    \centering
    \includegraphics[width=1.0\linewidth]{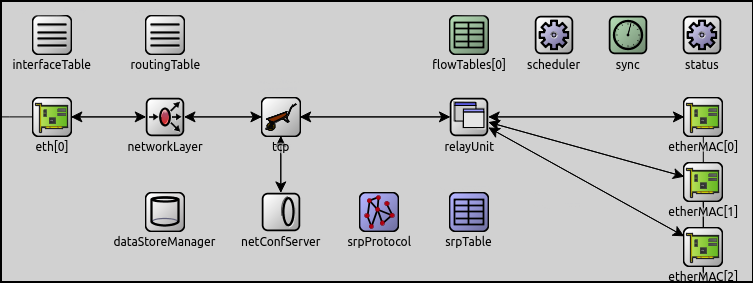}
    \caption{OMNeT++ Module of an OpenFlow and NetConf capable switch.}
    \label{fig:omnetswitch}
    \vspace{-15pt}
\end{figure}

Figure~\ref{fig:omnetswitch} shows the general module structure of a forwarding device that is capable of OpenFlow and NetConf.
The eth0 port connects the switch to the controller. 
The NetConf server module is connected to this port via TCP, but can also be connected to other protocols, e.g. for security reasons. 
It handles the NetConf protocol and extracts \acp{RPC} to forward them to the data store manager, which in turn applies the command on the proper data store and creates the response.
The OpenFlow protocol client is implemented in the relay unit and also connected via TCP.
Besides, the relay unit performs the flow table lookup in the flow tables and forwards incoming data packets to the correct Ethernet port.
For the etherMAC interfaces, we provide several real-time Ethernet port implementations that control the ingress and egress timings.
Some of them require the Time Synchronization and Scheduler modules, which also contribute a realistic device clock.
The stream reservation table and protocol modules are needed for \ac{AVB}/\ac{TSN} streams.
The switch implementation also provides the option to import or export a launch configuration for all modules and tables.    



The \ac{SDN} controller module is based on a standard host provided by the INET framework so that other applications can run smoothly.
The OpenFlow controller module from the OpenFlowOMNeTSuite and the NetConf client module connect to the forwarding devices via TCP. 
The NetConf client module handles connectivity to the forwarding devices and forwards \ac{RPC} requests and replies to the corresponding NetConf application or server. 
We include a base implementation for NetConf and OpenFlow applications as well as specific implementation for real-time controllers.
With this, it is possible to create controller apps that use NetConf and OpenFlow simultaneously.
The controller also provides the ability to import or export a launch configuration.


\section{Case Study} 
\label{sec:case_study}

\begin{figure}
    \centering
    \includegraphics[width=0.95\linewidth, trim= .8cm 1.3cm .55cm .85cm, clip=true]{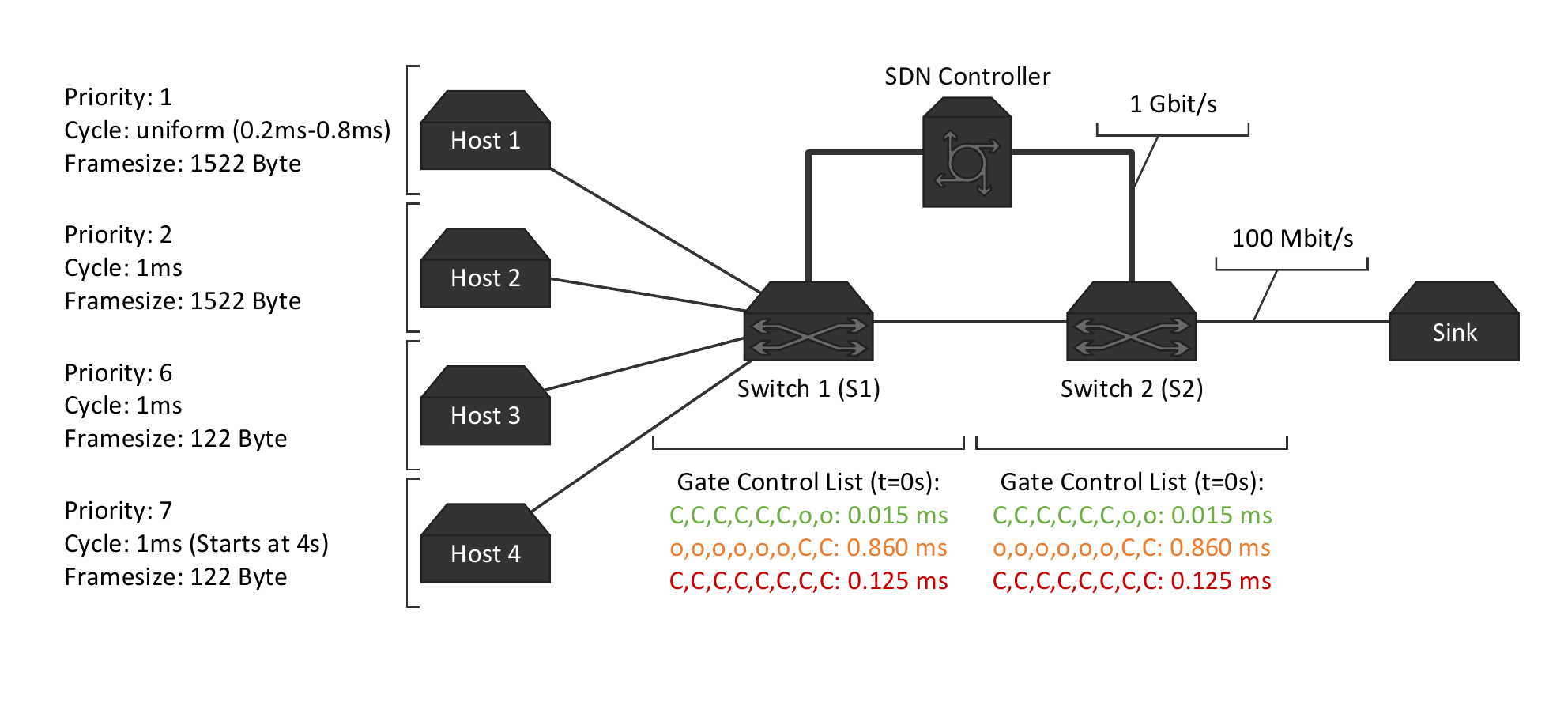}
    \vspace{-6pt}
    \caption{Network topology and device configurations at startup. }
    \label{fig:topology}
    \vspace{-8pt}
\end{figure}

Our results in previous work show that \ac{SDN} control overhead can be added to \ac{TSN} streams without a delay penalty, provided protocols are mapped properly~\cite{hmks-snsti-19}. 
In this work, we analyze the programmability of the \ac{GCL} in switches supporting IEEE 802.1Qbv. 
Since this article focuses on programming rather than calculating correct schedules, we have disabled clock jitter and time synchronization for all devices to simplify the simulation and calculation of schedules.

The network topology of our case study consists of four hosts sending messages to one sink via two switches in the presence of one \ac{SDN} controller as shown in Figure~\ref{fig:topology}. 
Each host sends one periodic IEEE 802.1Q message, so the traffic shaping is controlled by the switches in the network. 
These messages have a fixed priority, frame size, and cycle. 
As an exception, host 1 sends low priority messages with high frequency that varies according to gaussian normal distribution.
To introduce a change in high priority traffic, host 4 starts transmitting frames with the highest priority after 4 seconds.
The two switches are NetConf and OpenFlow capable. 
The output on each switch port is controlled by an 802.1Qbv module, which contains gates for each 802.1Q priority that can either be closed (C) or opened (O) at a certain point in time. 
This information is stored in the \ac{GCL} that can be programmed at run time using the NetConf protocol.
In this evaluation we use three phases for this list: 
A time slot $T_{red}$, closing all gates to ensure that the transmission of previous frames has ended before switching to the green phase; 
A time window $T_{green}$, for high priority traffic with priorities 6-7; 
And a window $T_{yellow}$, for all low priority traffic of priorities 0-5.

The cycle time of the schedule is equivalent to the largest cycle time of the high priority messages, which is \SI{1}{\ms} in our scenario. We calculated the phases and added a safety margin:
\begin{align}
    T_{red} 
    = T_{tx}^{Max Frame} + T_{margin} 
    = \frac{1,522\, Byte}{100\, Mbit/s} + 5\, \mu s 
    = 121,76\, \mu s + 5\, \mu s \approx 125\, \mu s 
    \\
    T_{green} 
    = T_{tx}^{Host 3} + T_{margin} 
    = \frac{122\, Byte}{100\, Mbit/s} + 5\, \mu s 
    = 9.76\, \mu s + 5\, \mu s \approx 15\, \mu s 
    \\
    T_{yellow} 
    = T_{cycle} - T_{green} - T_{red} 
    = 1\, ms - 15\, \mu s - 125\, \mu s = 860\, \mu s
\end{align}

Figure~\ref{fig:evaluation} is a graph showing the changes in end-to-end latency during $10s$ simulation time. 
These changes are introduced by updates to the configuration of the \ac{GCL} depicted in Table~\ref{tab:controllist} and changes in the traffic pattern.

At the beginning of the simulation all devices are configured as shown in Figure~\ref{fig:topology}. 
The high priority packets from host 3 to the sink have a constant delay of 1.03 ms (see Figure~\ref{fig:evaluation}). 
In the simulation, we can see that this delay is introduced at the second switch, where the packets miss their time slot because of the transmission time of $20\, \mu s$ over two links. 
To eliminate this delay, the controller updates the \ac{GCL} of both forwarding devices at $2s$ simulation time. 
The red phase is split into two parts, one at the beginning of the cycle and one at the end. 
The red phase at the beginning is exactly the transmission time of the high priority packet to the device. 
For switch 1 this is the transmission delay of $10\, \mu s$ from host 3 to switch 1. 
For switch 2 this phase has to be shifted another $10\, \mu s$ for the transmission from switch 1 to switch 2. 
This update reduces the end-to-end latency to the expected value of $30\, \mu s$.

At $4s$ simulation time, host 4 starts sending frames with the same size and cycle but a higher priority than those of host 3.
Therefore, the transmission time of high priority frames at the switches doubles.
The packets of host 4 have the expected end-to-end latency of $30\, \mu s$.
Those of host 3 stay in time for their slot at switch 1 because of the safety margin of $5\, \mu s$.
At switch 2, however, they miss their time slot and need to wait for one cycle.
To solve this time conflict, the controller updates the GCLs at $6\, s$ simulation time, but only for switch 1 to simulate a failure in the configuration of switch 2.
Since the update was unsuccessful on switch 2, the end-to-end latency remains unchanged.
Although the configuration of switch 2 was unsuccessful, the update of the configuration of switch 1 was still performed.
In very sensitive environments this can lead to dangerous behavior of the network.
Additional mechanisms are required to ensure that changes to the schedule are executed at the same time and are consistent across the network.
At $8s$ simulation time, the controller performs the update of switch 2.
This reduces the end-to-end latency of host 3 as expected.

\begin{figure}[t]
    \centering
    \vspace{-5pt}
    \begin{minipage}[b]{0.48\textwidth}
        \centering
        \begin{tikzpicture}
            \begin{axis}[width=.95\textwidth, height=.9\textwidth,
              scaled y ticks = false,
            x unit=\second,
            y unit=\milli\second,
            xlabel=Simulation time,
            ylabel=End-to-end latency,
            legend pos=north east,
            legend cell align={left},
                legend style={font=\small},
            xmin=0,
            xmax=10,
            xtick distance=2,
            xticklabel style={/pgf/number format/fixed,/pgf/number format/precision=3},
            ymin=0,
            ymax=1.4,
            ytick distance=.5,
            ]
            \addplot[lightcorered,
            very thick,
            ] table [x, y, col sep=comma] {data/app6clean.csv};
            \addplot[coredarkgray,
            very thick,
            ] table [x, y, col sep=comma] {data/app7clean.csv};
              \legend{Host 3 $\rightarrow$ Sink (PCP 6), Host 4 $\rightarrow$ Sink (PCP 7)}
            \end{axis}
        \end{tikzpicture}
        \vspace{-5pt}
        \captionof{figure}{The end-to-end latency of high priority data flows from the host to the sink. Host 4 starts sending after \SI{4}{\s} simulation time.\newline}
        \label{fig:evaluation}
        \vspace{-7pt}
    \end{minipage}
    \hspace{5pt}
    \begin{minipage}[b]{0.48\textwidth}
        \centering
        \hspace{-8pt}
        \begin{tabular}{c|l | p{.63\textwidth}}
            \textbf{t(\si{\s})}
            & \textbf{Dev}
            & \textbf{Gate Control List (\si{\micro s})}       
            \\\hline \hline
            &&\\[-7pt]
            \textbf{0-2}
            & S1
            & G:~15 ; Y:~860 ; R:~125 
            \\
            & S2
            & G:~15 ; Y:~860 ; R:~125
            \\\hline
            &\\[-7pt]
            \textbf{2-4}
            & S1                  
            & R:~10 ; G:~15 ; Y:~860 ; R:~115 
            \\    
            & S2
            & R:~20 ; G:~15 ; Y:~860 ; R:~105 
            \\\hline
            &\\[-7pt]
            \textbf{4-6}
            & S1              
            & R:~10 ; G:~15 ; Y:~860 ; R:~115 
            \\
            & S2
            & R:~20 ; G:~15 ; Y:~860 ; R:~105
            \\\hline
            &\\[-7pt]
            \textbf{6-8}
            & S1          
            & R:~10 ; G:~30 ; Y:~845 ; R:~115 
            \\    
            & S2
            & R:~20 ; G:~15 ; Y:~860 ; R:~105
            \\\hline
            &\\[-7pt]
            \textbf{8-10}
            & S1              
            & R:~10 ; G:~30 ; Y:~845 ; R:~115 
            \\    
            & S2
            & R:~20 ; G:~30 ; Y:~845 ; R:~105
        \end{tabular}
        \captionof{table}{Changes to the 802.1Qbv \acl{GCL} of the switches (S1 and S2) with the three phases green (G), yellow (Y) and red (R).}
        \label{tab:controllist}
        \vspace{4pt}
    \end{minipage}
    \vspace{-16pt}
\end{figure}
 
This initial evaluation shows how the programming of schedules in the network can increase the flexibility and performance of the system. 
On the other hand, it indicates that re-configuration of time slots should be used with care during runtime.
Configuring multiple devices might lead to an inconsistent state of the network and loss or delay of critical messages. 


\vspace{-5pt}
\section{Conclusions and  Future Work} 
\label{sec:conclusion_&_future_work}
In this paper, we shared our simulation model \ac{SDN4CoRE} and described in detail how to make real-time communication programmable with NetConf and  OpenFlow.
In a case study, we demonstrated the use of our model for testing and evaluating configuration mechanisms in real-time networks. 
An initial assessment indicates that re-configuration of time slots should be used with care during runtime, and additional mechanisms are required to ensure a consistent state across the network.

In future work, we plan to further extend the programmability of the CoRE simulation models, for example, with the recently published implementation of IEEE 802.1Qci modules. 
The evaluation of control plane concepts for automotive networks is one of our next tasks.

The \ac{SDN4CoRE} simulation model, including its original fork of the OpenFlowOMNeTSuite, as well as all other simulation models and analyses tools, are published as open-source on our GitHub page at \url{https://github.com/CoRE-RG/}.

\vspace{-5pt}
\section*{Acknowledgments}
\label{sect:acks}
This work is funded by the Federal Ministry of Education and Research of Germany (BMBF) within the SecVI project.

\label{sect:bib}
\vspace{-5pt}
\bibliographystyle{plain}
\bibliography{own,rfcs,HTML-Export/all_generated,bib/bibliography}



\end{document}